\documentclass[aps,prd,superscriptaddress,10pt,nofootinbib,preprintnumbers]{revtex4} 

\usepackage{amsfonts}
\usepackage{amssymb}
\usepackage{amsmath}

\begin{document}

\preprint{YITP-16-92, IPMU16-0106} 

\title{Stealth magnetic field in de Sitter spacetime}
\author{Shinji Mukohyama}
\affiliation{Center for Gravitational Physics, Yukawa Institute for Theoretical Physics, Kyoto University, 606-8502, Kyoto, Japan}
\affiliation{Kavli Institute for the Physics and Mathematics of the Universe (WPI), The University of Tokyo Institutes for Advanced Study, The University of Tokyo, Kashiwa, Chiba 277-8583, Japan}

\date{\today}

 \begin{abstract}
  In the context of a $U(1)$ gauge theory non-minimally coupled to scalar-tensor gravity, we find a cosmological attractor solution that represents a de Sitter universe with a homogeneous magnetic field. The solution fully takes into account backreaction of the magnetic field to the geometry and the scalar field. Such a solution is made possible by scaling-type global symmetry and fine-tuning of two parameters of the theory. If the fine-tuning is relaxed then the solution is deformed to an axisymmetric Bianchi type-I universe with constant curvature invariants, a homogeneous magnetic field and a homogeneous electric field. Implications to inflationary magnetogenesis are briefly discussed. 
 \end{abstract}

\maketitle

\section{Introduction}

Multi-frequency observations of $\gamma$-rays from distant blazars indicate the presence of magnetic fields in extragalactic void regions~\cite{Neronov:1900zz,Tavecchio:2010mk,Ando:2010rb,Dolag:2010ni,Essey:2010nd,Taylor:2011bn,Takahashi:2013lba,Chen:2014rsa}. While astrophysical mechanisms could be effective to address the origin of galactic magnetic fields, there is a lack of explaination for extragalactic magnetic fields with correlation length of order of Mpc via astrophysical mechanisms. For this reason, it is probably natural to seek a possible explanation in the early universe. However, in Maxwell theory of electromagnetism, the conformal symmetry prevents magnetic fields from being generated by the expansion of the universe. The standard Maxwell theory thus needs to be modified if one considers the expansion of the universe as the origin of the large-scale magnetic fields. While various theoretical models have been proposed (see \cite{Widrow:2002ud,Kandus:2010nw,Durrer:2013pga,Subramanian:2015lua} for review and \cite{Kobayashi:2014sga,Caprini:2014mja,Tasinato:2014fia,Fujita:2015iga,Domenech:2015zzi,Fujita:2016qab} for more recent proposals), there is no convincing scenario of magnetogenesis in the early universe so far. In this situation, it is important to explore various approaches toward our understanding of the origin of large-scale magnetic fields.

Typical problems that may arise in early universe scenarios of magnetogenesis are instability~\cite{Himmetoglu:2009qi}, backreaction and strong coupling~\cite{Demozzi:2009fu,Fujita:2012rb}. It would thus be of interest if there is a stable cosmological solution with magnetic fields that fully takes into account backreaction. In the present paper, we thus seek such a solution in a $U(1)$ gauge theory non-minimally coupled to scalar-tensor gravity. In an accompanying paper~\cite{paper2} it is shown that at subhorizon scales all coefficients of (time) kinetic terms and squared sound speeds of linear perturbations are positive, meaning that there is no instability faster than the cosmological scale, in a range of parameters.

The rest of the paper is organized as follows. In section~\ref{sec:model} we describe the theory in which we seek a cosmological solution with a magnetic field. In section~\ref{sec:solution} we seek a fixed-point solution of the system. Generically, the solution represents an axisymmetric Bianchi type-I universe with constant curvature invariants, a homogeneous magnetic field and a homogeneous electric field. Upon fine-tuning two parameters in the theory, one obtains a de Sitter universe with a homogeneous magnetic field but without electric field. In section~\ref{sec:attractor} we then seek the condition under which the de Sitter fixed-point solution is an attractor of the system. Section~\ref{sec:summary} is devoted to a summary of the paper and some discussions. Throughout this paper (except in the discussion about implications of the model to magnetogenesis in Section~\ref{sec:summary}) we adopt the unit in which $M_{\rm Pl}=1$.

\section{Model description}
\label{sec:model}

The model consists of a metric $g_{\mu\nu}$, a $U(1)$ gauge field $A_{\mu}$ and a scalar field $\phi$. We suppose that the action of the system is invariant under the $U(1)$ gauge transformation 
\begin{equation}
 A_{\mu} \to A_{\mu} + \partial_{\mu} \lambda\,,  \label{eqn:gaugesym}
\end{equation}
as well as the scaling-type global symmetry transformation
\begin{equation}
 \phi \to \phi + \phi_0\,, \quad A_{\mu} \to e^{-\phi_0} A_{\mu}\,.
  \label{eqn:globalsym}
\end{equation}
Here, the mass dimension of $\phi$ is zero, $\lambda$ is an arbitrary function and $\phi_0$ is an arbitrary constant. For simplicity we demand that the equations of motion for $g_{\mu\nu}$, $A_{\mu}$ and $\phi$ are up to second-order differential equations.

In order to construct the action of the system, it is convenient to define the following tensors invariant under both the $U(1)$ gauge transformation (\ref{eqn:gaugesym}) and the global symmetry transformation (\ref{eqn:globalsym}):
\begin{equation}
 \mathcal{F}_{\mu\nu} \equiv e^{\phi}F_{\mu\nu}\,,
  \quad
  \tilde{\mathcal{F}}^{\mu\nu} \equiv e^{\phi}\tilde{F}^{\mu\nu}\,,
\end{equation}
where
\begin{equation}
 F_{\mu\nu} \equiv \partial_{\mu}A_{\nu} - \partial_{\nu}A_{\mu}\,, \quad
  \tilde{F}^{\mu\nu} \equiv \frac{1}{2}\epsilon^{\mu\nu\rho\sigma}F_{\rho\sigma}\,,
\end{equation}
are the field strength and its Hodge dual, and the Levi-Civita tensor is normalized as $\epsilon^{0123}=-1/\sqrt{-g}$. The building blocks of the action are then
\begin{equation}
 g_{\mu\nu}\,, \ g^{\mu\nu}\,, \ \nabla_{\mu}\,, \ R^{\mu}_{\ \nu\rho\sigma}\,, \ \mathcal{F}_{\mu\nu}\,, \  \tilde{\mathcal{F}}^{\mu\nu}\,, \ \partial_{\mu}\phi\,, \ \cdots\,,
\end{equation}
where $g^{\mu\nu}$ and $R^{\mu}_{\ \nu\rho\sigma}$ are the inverse and the Riemann curvature of $g_{\mu\nu}$, and $\nabla_{\mu}$  is the covariant derivative compatible with $g_{\mu\nu}$. It is easy to show by the same logic as \cite{Fleury:2014qfa} that any scalar function made of $g_{\mu\nu}$,  $g^{\mu\nu}$, $\partial_{\mu}\phi$, $\mathcal{F}_{\mu\nu}$, $\tilde{\mathcal{F}}^{\mu\nu}$ without derivatives acted on them can be written as a function of the following four scalar combinations
\begin{equation}
 X \equiv  -\frac{1}{2}g^{\mu\nu}\partial_{\mu}\phi\partial_{\nu}\phi\,,\quad
 W \equiv  -\frac{1}{4}\mathcal{F}_{\mu\nu}\mathcal{F}^{\mu\nu}\,,\quad
 Y \equiv  \mathcal{F}_{\mu\nu}\tilde{\mathcal{F}}^{\mu\nu}\,,\quad
 Z \equiv  \mathcal{F}^{\rho\mu}\mathcal{F}_{\rho}^{\ \nu}\partial_{\mu}\phi\partial_{\nu}\phi\,. 
\end{equation} 
It is also easy to modify the Horndeski's non-minimal coupling of a $U(1)$ gauge field to the Riemann tensor~\cite{Horndeski:1976gi} in a way that renders it consistent with the global symmetry (\ref{eqn:globalsym}). The resulting invariant non-minimal coupling is 
\begin{equation}
 L_{\rm H} = \xi\tilde{\mathcal{F}}^{\mu\nu}\tilde{\mathcal{F}}^{\rho\sigma}R_{\mu\nu\rho\sigma}\,, 
\end{equation}
where $\xi$ is an arbitrary constant. We can also add shift-symmetric Horndeski terms~\cite{Horndeski:1974wa,Deffayet:2011gz} for the scalar field $\phi$. We thus end up with the following action
\begin{equation}
 I = \int d^4x \sqrt{-g}
  \left[ L + L_3 + L_4 + L_5 + L_{\rm H}\right]\,, \label{eqn:action}
\end{equation}
where $L=L(X,W,Y,Z)$ is an arbitrary function of $(X,W,Y,Z)$, and 
\begin{eqnarray}
 L_3 & = & -G_3(X)\Box\phi\,, \nonumber\\
 L_4 & = & G_4(X) R + G_{4X}(X)\left[(\Box\phi)^2-(\nabla^{\mu}\nabla_{\nu}\phi)(\nabla^{\nu}\nabla_{\mu}\phi)\right]\,,\nonumber\\
 L_5 & = & G_5(X)G^{\mu\nu}\nabla_{\mu}\nabla_{\nu}\phi 
  - \frac{1}{6}G_{5X}(X)
  \left[ (\Box\phi)^3 - 3(\Box\phi)(\nabla^{\mu}\nabla_{\nu}\phi)(\nabla^{\nu}\nabla_{\mu}\phi) + 2(\nabla^{\mu}\nabla_{\nu}\phi)(\nabla^{\nu}\nabla_{\rho}\phi)(\nabla^{\rho}\nabla_{\mu}\phi) \right]\,,
\end{eqnarray}
are general shift-symmetric Horndeski terms for $\phi$. Here, $G_3(X)$, $G_4(X)$ and $G_5(X)$ are arbitrary functions of $X$ and the subscript $X$ denotes derivative with respect to $X$.

\section{Fixed-point solution}
\label{sec:solution}

\subsection{Ansatz}

We consider a homogeneous scalar field,
\begin{equation}
 \phi = \phi(t)\,.
\end{equation}
in an axisymmetric Bianchi type-I spacetime, 
\begin{equation}
 g_{\mu\nu}dx^{\mu}dx^{\nu}  = \eta_{ab}e^a e^b
  = -N(t)^2dt^2 + a(t)^2
  \left[ e^{4\sigma(t)}dx^2 + e^{-2\sigma(t)}(dy^2+dz^2)\right]\,,
\end{equation}
where $\eta_{ab}={\rm diag} (-1,1,1,1)$ and
\begin{equation}
 e^0 = N(t)dt\,, \quad e^1 = a(t) e^{2\sigma(t)}dx\,, \quad
  e^2 = a(t) e^{-\sigma(t)}dy\,, \quad   e^3 = a(t) e^{-\sigma(t)}dz\,.
  \label{eqn:def-e^a}
\end{equation}
As for the $U(1)$ gauge field, we consider the following ansatz,
\begin{equation}
 A_t = 0\,,\ A_x = \int^t \frac{N(t')e^{4\sigma(t')}}{a(t')}E(t')dt'\,,\
  A_y = \frac{1}{2}Bz\,,\   A_z = -\frac{1}{2}By\,, \label{eqn:A-ansatz}
\end{equation}
where $E(t)$ is a function of $t$ and $B$ is a constant, so that
\begin{equation}
 \frac{1}{2}{\cal F}_{\mu\nu}dx^{\mu}\wedge dx^{\nu} =
  E \chi e^0\wedge e^1 - B\chi e^2\wedge e^3\,,
\end{equation}
and that 
\begin{equation}
 X = \frac{\dot{\phi}^2}{2N^2}\,,\ W =\frac{1}{2}(E^2-B^2)\chi^2\,,\ Y = 4EB\chi^2 \,,\
  Z =  2E^2\chi^2 X\,. \label{eqn:XWYZ}
\end{equation}
Here, an over-dot represents derivative w.r.t. $t$ and 
\begin{equation}
 \chi \equiv \frac{e^{\phi}e^{2\sigma}}{a^2}\,. \label{eqn:def-chi}
\end{equation}
It is straightforward to calculate the equations of motion for $g_{\mu\nu}$, $A_{\mu}$ and $\phi$. The independent equations are the four equations shown in Appendix~\ref{app:eom}, where
\begin{equation}
 H \equiv \frac{\dot{a}}{Na}\,, \quad  \Sigma \equiv \frac{\dot{\sigma}}{N}\,.
  \label{eqn:def-HSigma}
\end{equation}

\subsection{Anisotropic fixed-point solution}

We would like to find solutions for which the scalar invariants ($X$, $W$, $Y$, $Z$) shown in (\ref{eqn:XWYZ}) are constant. (We have already assumed that $B$ is constant in (\ref{eqn:A-ansatz}).) We further demand that scalar invariants made of the metric and its curvature are also constant. These demands are fulfilled if and only if $\dot{\phi}/N$, $E$ ($=E_0$), $\chi$, $H$ ($=H_0$) and $\Sigma$ ($=\Sigma_0$) are constant, where $\chi$, $H$ and $\Sigma$ are defined in (\ref{eqn:def-chi}) and (\ref{eqn:def-HSigma}). Hereafter, by overall re-scaling of spatial coordinates, we set
\begin{equation}
 \chi = 1\,.
\end{equation}
Under these requirements and the overall normalization of spatial coordinates, the solution is characterized by the four parameters ($H_0$, $\Sigma_0$, $E_0$, $B$). (For example, the constancy of $\chi$ implies that $X=2(H_0-\Sigma_0)^2$.) Since the number of independent equations of motion is also four, a generic choice of $L(X,W,Y,Z)$, $G_{3,4,5}(X)$ and $\xi$ in the action allows for such a solution.

Since we are interested in an expanding universe, we suppose that $H_0$ is positive. For later convenience, we introduce three dimensionless quantities
\begin{equation}
 s \equiv \frac{\Sigma_0}{H_0}\,, \quad e \equiv \frac{E_0}{H_0}\,, \quad b \equiv \frac{B}{H_0}\,, 
\end{equation}
and consider ($H_0$, $s$, $e$, $b$) as four independent parameters characterizing the solution.  While the expansion rate $H_0$ sets the overall energy scale of the system, $s$, $e$ and $b$ correspond to the dimensionless sizes of the anisotropy, the electric field and the magnetic field, respectively.

Upon setting
\begin{equation}
 N = 1\,, \ a = e^{H_0t}\,,\ \sigma=sH_0t\,,\ \phi = 2(1-s)H_0t\,, \ E=eH_0\,,\ B = bH_0\,,
\end{equation}
where $H_0$, $s$, $e$ and $b$ are constants, the scalar invariants ($X$, $W$, $Y$, $Z$) are expressed as 
\begin{equation}
 X = 2(1-s)^2H_0^2\,,\ W = \frac{1}{2}(e^2-b^2)H_0^2\,,\ Y = 4ebH_0^2\,,\ Z = 4(1-s)^2e^2H_0^4\,.
\end{equation}
As a result, the equations of motion greatly simplify as
\begin{eqnarray}
 0 & = & 8(1-s)^2(L_Z+\xi)eH_0^2 + eL_W + 4bL_Y\,,\nonumber\\
 0 & = & 16(1-s)^6H_0^6G_{5X}
 + 4\left[(2L_Z+\xi)e^2+4\xi b^2+6(1-s)^2G_{4X}\right](1-s)^2H_0^4 \nonumber\\
 & & + \left[L_We^2+4L_Yeb-6G_4(1-s)^2\right]H_0^2 - L\,,\nonumber\\
 0 & = & 8G_{5XX}(1+2s)(1-s)^6H_0^6
   + 6(1-s)^4\left[4(1+s)G_{4XX}+(1+4s)G_{5X}\right]H_0^4 
  \nonumber\\
 & &
  + 2\left\{(1-s)(L_Z+\xi)e^2-(1-4s)\xi b^2 + 3(1-s)^2[G_{3X}+(1+3s)G_{4X}]\right\}H_0^2
  -3sG_4 + (1-s)L_X \,,\nonumber\\
 0 & = & 72G_{5X}(1-s)^4sH_0^4 + 4\left\{ (1-s)[(1+2s)\xi-2(1-s)L_Z]e^2 \right. \nonumber\\
& &  \left. -\xi(5-4s+8s^2)b^2 + 18s(1-s)^2G_{4X} \right\}H_0^2 
 -(e^2+b^2)L_W - 18sG_4\,, \label{eqn-eom-aniso}
\end{eqnarray}
where subscripts $X$, $W$, $Y$ and $Z$ represent derivatives with respect to them. These four equations are algebraic equations for four unknown constants ($H_0$, $s$, $e$, $b$). Once functions $L(X,W,Y,Z)$, $G_3(X)$, $G_4(X)$, $G_5(X)$ and a constant $\xi$ are specified, one can in principle solve these four algebraic equations for four constants ($H_0$, $s$, $e$, $b$).

\subsection{de Sitter fixed-point solution}

If we fine-tune the action of the system so that $s\to 0$, then we obtain a de Sitter solution. By setting $s=0$, the equations of motion (\ref{eqn-eom-aniso}) reduce to 
\begin{eqnarray}
0 & = & 8(L_Z+\xi)eH_0^2 + eL_W + 4bL_Y\,,\nonumber\\
0 & = & 16H_0^6G_{5X} 
 + 4\left[(2L_Z+\xi)e^2+4\xi b^2+6G_{4X}\right]H_0^4 + (L_We^2+4L_Yeb-6G_4)H_0^2 - L\,,\nonumber\\
0 & = &  8G_{5XX}H_0^6 + 6(4G_{4XX}+G_{5X})H_0^4 + 2\left[ (L_Z+\xi)e^2-\xi b^2 + 3(G_{3X}+G_{4X})\right]H_0^2+ L_X\,,\nonumber\\
0 & = &  4\left[(\xi-2L_Z)e^2 -5\xi b^2 \right]H_0^2 -(e^2+b^2)L_W\,.
\end{eqnarray}

\subsection{de Sitter fixed-point solution without electric field}
\label{subsec:dSnoE}

If we further fine-tune the action of the system so that $s\to 0$ and $e\to 0$ simultaneously, then we obtain a de Sitter solution without electric field. By setting $s=e=0$ and assuming that $b\ne 0$, the equations of motion (\ref{eqn-eom-aniso}) reduce to 
\begin{eqnarray}
0 & = &   L_Y\,,\nonumber\\
0 & = &   16G_{5X}H_0^6 + 8(2\xi b^2+3G_{4X})H_0^4 -6G_4H_0^2 - L\,,\nonumber\\
0 & = &   8G_{5XX}H_0^6 + 6(4G_{4XX}+G_{5X})H_0^4 + 2\left[ -\xi b^2 + 3(G_{3X}+G_{4X})\right]H_0^2+ L_X\,,\nonumber\\
0 & = &   20\xi H_0^2 + L_W\,. \label{eqn:dSnoE}
\end{eqnarray}
The ansatz under consideration is parametrized by two constants ($H_0$, $b$). We thus need to fine-tune two parameters in the action to make this set of four algebraic equations to be solvable w.r.t. ($H_0$, $b$). One of the two fine-tunings can be easily achieved if we demand that the function $L$ is even w.r.t. $Y$. Actually, in this case the first equation, $L_Y=0$, is automatically satisfied.

\section{Attractor behavior}
\label{sec:attractor}

In this section we seek the condition under which the de Sitter fixed-point solution without electric field (and thus with $Y=0$) introduced in subsection~\ref{subsec:dSnoE} is a local attractor of the system. For simplicity, we assume that the function $L$ is even w.r.t. $Y$ so that odd-order derivatives of $L$ w.r.t. $Y$ vanish on any backgrounds with $Y=0$. In this case the first equation in (\ref{eqn:dSnoE}) is trivial. Furthermore, we fine-tune one parameter in the action so that the set of the second, third and forth algebraic equations in (\ref{eqn:dSnoE}) is solvable w.r.t. the two constants ($H_0$, $b$).

We set 
\begin{equation}
N = 1\,, \quad
  H(t) = H_0(1+\epsilon h_1(t))\,, \quad 
  \Sigma(t) = \epsilon H_0s_1(t)\,, \quad
 \chi(t) = 1 + \epsilon\chi_1(t)\,, \quad
  E(t) = \epsilon e_1(t)H_0\,,\quad
  B = b H_0\,,
\end{equation}
and expand the four equations of motion summarized in Appendix~\ref{app:eom} w.r.t. $\epsilon$. At the order $O(\epsilon^0)$, we obtain the second, third and forth equations in (\ref{eqn:dSnoE}). The first equation in (\ref{eqn:dSnoE}) is trivially satisfied under the above mentioned assumption that $L$ be even w.r.t. $Y$. At the order $O(\epsilon)$, we obtain 
\begin{equation}
 \frac{1}{H_0}\frac{d}{dt}
  \left(
   \begin{array}{c}
    \chi_1\\
    h_1\\
    s_1\\
    e_1
   \end{array}
			\right) = {\bf R} 
   \left(
   \begin{array}{c}
    \chi_1\\
    h_1\\
    s_1\\
    e_1
   \end{array}
			      \right)\,,
\end{equation} 
where ${\bf R}$ is a $4\times 4$ matrix whose components are independent of the perturbations ($\chi_1$, $h_1$, $s_1$, $e_1$). The four eigenvalues of ${\bf R}$ are solutions of the following forth-order algebraic equation for $\lambda$: 
\begin{equation}
 0 = \det \left[\lambda \bf{1}_4 - {\bf R} \right]
= (\lambda + 4)(\lambda + 3)\left(\lambda^2 + 3\lambda + \frac{\mathcal{A}}{\mathcal{N}}\right)\,,
\end{equation}
where $\bf{1}_4$ is the $4\times 4$ identity matrix, and
\begin{eqnarray}
 \mathcal{N} & = & 2\zeta_3g_h(\zeta_3-8\zeta_1)b^2+\zeta_1(\zeta_1\zeta_2+3\zeta_3^2)\,,\nonumber\\
 \mathcal{A} & =  & 56b^6g_h^3-4(9\zeta_1+\zeta_2+15\zeta_3)g_h^2b^4-2g_h\zeta_4(\zeta_1-\zeta_3)b^3\nonumber\\
 & & 
 +\left[6(-\zeta_1^2+\zeta_1\zeta_2+2\zeta_1\zeta_3+2\zeta_3^2)g_h+\zeta_5(\zeta_1-\zeta_3)^2\right]b^2+\frac{3}{2}\zeta_1\zeta_4(\zeta_1-\zeta_3)b \,.
\end{eqnarray}
Here, $\zeta_i$ ($i=1,\cdots 5$) and $g_h$ are constants defined by
\begin{eqnarray}
 \zeta_1 & = & 2b^2g_h+g_4-4g_{4x}-4g_{5x}\,,\nonumber\\
 \zeta_2 & = & 2b^2g_h+6g_{3x}+24g_{3xx}+72g_{4xx}+96g_{4xxx}+6g_{5x}+48g_{5xx}+32g_{5xxx}+4l_{xx}  \,,\nonumber\\
 \zeta_3 & = & 4b^2g_h+2g_{3x}+4g_{4x}+16g_{4xx}+6g_{5x}+8g_{5xx}\,,\nonumber\\
 \zeta_4 & = & -4(g_h+l_{xw})b  \,,\nonumber\\
 \zeta_5 & = & -b^2l_{ww}-12g_h\,,\nonumber\\
  g_h & = & \xi\frac{H_0^2}{M_{\rm Pl}^2}\,, 
\end{eqnarray}
and
\begin{eqnarray}
 & &
  L_{XX} = l_{xx}\frac{M_{\rm Pl}^2}{H_0^2}\,, \quad
  L_{XW} = l_{xw}\frac{M_{\rm Pl}^2}{H_0^2}\,, \quad
  L_{WW} = l_{ww}\frac{M_{\rm Pl}^2}{H_0^2}\,, \quad
  G_{3X} = g_{3x}\frac{M_{\rm Pl}^2}{H_0^2}\,, \quad
  G_{3XX} = g_{3xx}\frac{M_{\rm Pl}^2}{H_0^4}\,, \nonumber\\
 & & 
  G_4 = g_4M_{\rm Pl}^2\,, \quad
  G_{4X} = g_{4x}\frac{M_{\rm Pl}^2}{H_0^2}\,, \quad
  G_{4XX} = g_{4xx}\frac{M_{\rm Pl}^2}{H_0^4}\,, \quad
  G_{4XXX} = g_{4xxx}\frac{M_{\rm Pl}^2}{H_0^6}\,,\nonumber\\
 & & 
  G_{5X} = g_{5x}\frac{M_{\rm Pl}^2}{H_0^4}\,, \quad
  G_{5XX} = g_{5xx}\frac{M_{\rm Pl}^2}{H_0^6}\,, \quad
  G_{5XXX} = g_{5xxx}\frac{M_{\rm Pl}^2}{H_0^8}\,. \label{eqn:dimensionlessparameters}
\end{eqnarray}
It is understood that the left hand sides of (\ref{eqn:dimensionlessparameters}) are evaluated at the fixed-point solution under consideration and thus are constant.

The necessary and sufficient condition for the de Sitter fixed-point solution without electric field to be an attractor of the system is that the real parts of the four eigenvalues of ${\bf R}$ be negative. On the other hand, by analyzing inhomogeneous (i.e. ($x$, $y$, $z$)-dependent) linear perturbations around the solution, one can show that the absence of ghost degrees of freedom requires~\cite{paper2}
\begin{equation}
 \mathcal{N}>0\,. \label{eqn:N>0}
\end{equation}
Under the condition (\ref{eqn:N>0}), the attractor condition is equivalent to 
\begin{equation}
 \mathcal{A} > 0\,. 
\end{equation}

\section{Summary and discussion}
\label{sec:summary}

In the context of a $U(1)$ gauge theory non-minimally coupled to scalar-tensor gravity, we have found a cosmological attractor solution in which a de Sitter universe supports a homogeneous magnetic field. The solution fully takes into account backreaction of the magnetic field to the geometry and the scalar field. Such a solution is made possible by scaling-type global symmetry and fine-tuning of two parameters of the theory. If the fine-tuning is relaxed then the solution is deformed to an axisymmetric Bianchi type-I universe with constant curvature invariants, a homogeneous magnetic field and a homogeneous electric field.

The system described by the action (\ref{eqn:action}) respects the diffeomorphism invariance and the $U(1)$ gauge symmetry, and its equations motion are up to second-order differential equations. Therefore the system contains five physical degrees of freedom: two from $g_{\mu\nu}$, two from $A_{\mu}$ and one from $\phi$. It is straightforward (though complicated) to analyze general inhomogeneous, i.e. $(x,y,z)$-dependent, linear perturbations around the de Sitter attractor solution without electric field. After fine-tuning two parameters as prescribed in the present paper, one can still find a range of parameters in which all coefficients of (time) kinetic terms and squared sound speeds of the five degrees of freedom are positive at subhorizon scales~\cite{paper2}, meaning that there is no instability faster than the cosmological expansion. (On the other hand, one does not necessarily need to require the positivity of coefficients of (time) kinetic terms and squared sound speeds in the infrared, i.e. at superhorizon scales~\cite{Gumrukcuoglu:2016jbh}.)

The de Sitter attractor solution with a stealth magnetic field that we have found in the present paper may be useful to address the origin of large-scale magnetic fields in the universe. For example, suppose that the scaling-type global symmetry (\ref{eqn:globalsym}) is maintained for small $\phi$ (possibly including the limit $\phi\to -\infty$) but that for large values of $\phi$ the global symmetry is broken and $\phi$ acquires a potential with a minimum. By arranging the system so that the symmetry breaking occurs after inflation, the homogeneous magnetic field is maintained during inflation but the system behaves as the standard Einstein-Maxwell system at late time. In order to suppress the statistical anisotropy and non-Gaussianity of curvature perturbations, one probably needs to introduce another field (or other fields) as inflaton or/and curvaton, instead of considering $\phi$ itself as the main source of curvature perturbations. In this case the exact attractor solution found in the present paper provides a background (quasi) de Sitter expansion on which a field responsible for the generation of curvature perturbations safely generates adiabatic and essentially statistically isotropic, Gaussian fluctuations.

Here, for simplicity let us suppose that the symmetry breaking and thus the stabilization of $\phi$ occur immediately after inflation. Denoting the value of $\phi$ at (and after) the end of inflation as $\phi_f$ and recovering $M_{\rm Pl}$ (which we set to unity for simplicity in the main body of the present paper), the amplitude of the magnetic field at the end of inflation is
\begin{equation}
 \mathcal{B}_f = e^{-\phi_f}M_{\rm Pl}H_0 |b|\,.
\end{equation}
After inflation and the stabilization of $\phi$ to $\phi_f$, the magnetic field decays adiabatically. Its present value is thus
\begin{equation}
 \mathcal{B}_{\rm today} = \mathcal{B}_f\left(\frac{a_f}{a_{\rm today}}\right)^2\,,
\end{equation}
where $a_f$ and $a_{\rm today}$ are the scale factor at the end of inflation and its present value, respectively. Supposing that the universe is dominated by inflaton oscillation between the end of inflation ($a=a_f$) and the onset of the radiation dominated epoch ($a=a_R$), the scale factor today is estimated by the entropy conservation as
\begin{equation}
 a_{\rm today} \simeq a_f\ g^{1/12}\frac{\sqrt{M_{Pl}H_0}}{T_{\rm today}}
  \left(\frac{a_R}{a_f}\right)^{1/4}\,,
\end{equation}
where $T_{\rm today}$ is the photon temperature today and $g$ is the number of relativistic degrees of freedom that eventually inject entropy to photons. With the instantaneous reheating approximation ($a_R=a_f$) and supposing that $g^{1/6}=O(1)$, we obtain
\begin{equation}
 \mathcal{B}_{\rm today} \simeq e^{-\phi_f}|b| T_{\rm today}^2 \simeq e^{-\phi_f}|b|\times 10^{-6} G\,.
\end{equation}
Intriguingly, this is independent of the scale of inflation. The current upper bound on the large scale magnetic field is roughly $10^{-9}G$~\cite{Shaw:2010ea}. On the other hand, the lower bound from the blazar observations is roughly $10^{-15}G$~\cite{Neronov:1900zz,Tavecchio:2010mk,Ando:2010rb,Dolag:2010ni,Essey:2010nd,Taylor:2011bn,Takahashi:2013lba,Chen:2014rsa}. Putting them together, we obtain the observational constraint on the combination $e^{-\phi_f}|b|$ as
\begin{equation}
10^{-9}\lesssim e^{-\phi_f}|b| \lesssim 10^{-3}\,. \label{eqn:constraint}
\end{equation}
It seems relatively easy to satisfy (\ref{eqn:constraint}), thanks to the exponential dependence on $\phi_f$.

There have been a number of severe constraints on inflationary magnetogenesis scenarios in the literature. As far as the author knows, the model presented in this paper can evade all of them, provided that (\ref{eqn:constraint}) is fulfilled. For example, ref.~\cite{Fujita:2014sna} obtained a strong constraint from the large energy-momentum tensor of electromagnetic field in the epoch between the horizon exit of the scale of interest and the end of inflation. This constraint does not apply to our model with $s\simeq 0$ since the background energy-momentum tensor is essentially proportional to the background metric and thus is indistinguishable from an effective cosmological constant. Ref.~\cite{Green:2015fss} obtained two different constraints, one classical and the other quantum. The classical constraint is again due to a large energy-momentum tensor of electromagnetic field during inflation and thus does not apply to our model. The quantum constraint does not apply either since our model contains magnetic field with a finite amplitude already at the level of a classical background solution that fully takes into account backreaction.

In our model the de Sitter expansion during inflation is realized by means of fine-tuning of a parameter in the action. If we de-tune it then the expansion in the attractor solution becomes anisotropic during inflation while homogeneity is still maintained. One might thus worry about the fact that the cosmic microwave background (CMB) observation disfavors Bianchi universes and the shear is severely constrained by the Planck data~\cite{Saadeh:2016sak}. However, those observational constraints are written in terms of the present value of the shear. Since the shear decays as $\propto 1/a^3$ and thus the corresponding energy density rather quickly decays as $\propto 1/a^6$ by the cosmic expansion, those apparently strong constraints become rather weak if written in terms of the value of the shear at the end of inflation. Indeed, $s=O(1)$ can easily satisfy those observational constraints presented in \cite{Saadeh:2016sak}. For this reason the CMB observation so far does not put a strong constraint on the model considered in the present paper, as far as the geometric contribution to the statistical anisotropy is concerned. On the other hand, the primordial stochastic contribution to the statistical anisotropy is expected to be induced by a non-vanishing $s$. It is worthwhile analyzing such a contribution in detail. The bottom line is that the present model should be totally consistent with observational data as far as $|s|$ is small enough.

As explained above, the homogeneous background magnetic field in our model may be the origin of the large scale magnetic field in the void region and seems consistent with all observational data so far, provided that the condition (\ref{eqn:constraint}) holds. At smaller scales, the same homogeneous magnetic field can act as the seed for the dynamo and compression amplification mechanisms in galaxies and clusters of galaxies. In our model we thus do not need to introduce inhomogeneneities to the magnetic field during inflation to explain the magnetic field in the universe today at various scales. It is nonetheless interesting to investigate what happens if inhomogeneous fluctuations of the magnetic field are super-imposed on top of the homogeneous background magnetic field. While the homogeneous background magnetic field leads to a non-vanishing Alfv\'{e}n velocity $v_A\sim 4\times 10^{-4}(B_{\rm today}/10^{-9}G)$, the power of the inhomogeneous perturbation of the magnetic field could be either constrained by the CMB data or considered as a possible explanation for the physical basis for some of the CMB anomalies~\cite{Durrer:1998ya,Ade:2013nlj}. For example, the Planck data constrains the power of vector perturbation $A_v$ at the pivot scale $0.05/Mpc$ as $A_v v_A^2\lesssim 10^{-11}$ (see Appendix of \cite{Ade:2013nlj}). This translates to a constraint on the combination $A_v e^{-2\phi_f}b^2$ as
\begin{equation}
 A_v e^{-2\phi_f}b^2 \lesssim 10^{-8}\,. 
\end{equation}
Considering the bound (\ref{eqn:constraint}), this is not a strong restriction on our model. It is nonetheless intriguing to push forward this kind of constraints/possibilities.

It is known that MHD turbulence can be developed by coupling between a magnetic field and the primordial plasma. In the case of a primordial stochastic magnetic field, results of MHD simulations indicate that the spectrum of the magnetic field remains unchanged on large scales~\cite{Kahniashvili:2012vt}. Thus the homogeneous magnetic field is also expected to survive the MHD turbulence in our universe. It is desirable to confirm this explicitly by MHD simulations. If confirmed, our scenario has a potential to explain the magnetic field in our universe at all scales.

In summary the exact attractor solution found in the present paper provides a basis for a new type of inflationary magnetogenesis by which the origin of magnetic fields in our universe at all scales may be explained by a homogeneous magnetic field. It is worthwhile to study this scenario of magnetogenesis in more detail.

\begin{acknowledgments}
This work was supported in part by JSPS KAKENHI Grant Number 24540256 and World Premier International Research Center Initiative (WPI), MEXT, Japan.
\end{acknowledgments}

\appendix

\section{Equations of motion}
\label{app:eom}

The four independent equations of motion are
\begin{eqnarray}
 0 & = & L - L_X\left(\frac{\dot{\phi}}{N}\right)^2 - E^2 L_W\chi^2 - 4BE L_Y\chi^2 - 4E^2L_Z \chi^2\left(\frac{\dot{\phi}}{N}\right)^2  - 3HG_{3X}\left(\frac{\dot{\phi}}{N}\right)^3
  \nonumber\\
 & & + 6(H^2-\Sigma^2)\left[G_4-2G_{4X}\left(\frac{\dot{\phi}}{N}\right)^2 - G_{4XX}\left(\frac{\dot{\phi}}{N}\right)^4\right] - (H-\Sigma)^2(H+2\Sigma)\left(\frac{\dot{\phi}}{N}\right)^3 \left[5G_{5X} + G_{5XX}\left(\frac{\dot{\phi}}{N}\right)^2 \right] \nonumber\\
 & & + 4\xi\chi^2\left[ 2(H+2\Sigma)\left(H-\Sigma-\frac{\dot{\phi}}{N}\right)B^2 - 3(H-\Sigma)^2E^2 \right]\, ,
\end{eqnarray}
\begin{eqnarray}
 0 & = &
  \Bigg\{ 4L_{ZZ}E^2\chi^2\left(\frac{\dot{\phi}}{N}\right)^4
   + 2\left[2(4L_{YZ}B+L_{ZW}E)E\chi^2+L_Z\right]\left(\frac{\dot{\phi}}{N}\right)^2
   \nonumber\\
 & & 
   + (16L_{YY}B^2+8L_{WY}BE+L_{WW}E^2)\chi^2 + L_W + 8(H-\Sigma)^2\xi \Bigg\}
  \frac{\dot{E}}{N} \nonumber\\
 & & 
  \left[ 2E(2E^2\chi^2L_{ZZ}+L_{XZ})\left(\frac{\dot{\phi}}{N}\right)^2
   + 2E^2\chi^2(4BL_{YZ} + EL_{ZW})  + 4BL_{XY} + E(L_{XW} + 4L_Z) \right]\frac{\dot{\phi}}{N}\frac{1}{N}\frac{d}{dt}\left(\frac{\dot{\phi}}{N}\right) \nonumber\\
 & & 
+16E\xi (H-\Sigma)\left(\frac{\dot{H}}{N}-\frac{\dot{\Sigma}}{N}\right) + 2(EL_W+4BL_Y)\frac{\dot{\phi}}{N} + 4EL_Z\left(\frac{\dot{\phi}}{N}\right)^3 \nonumber\\
 & &
+ \left(2H-2\Sigma-\frac{\dot{\phi}}{N}\right)\chi^2\left[E(B^2-E^2)L_{WW} - 32EB^2L_{YY}  -4E^3\left(\frac{\dot{\phi}}{N}\right)^4 L_{ZZ} + 4B(B^2-3E^2)L_{WY} \right. \nonumber\\
 & & \left. -24E^2B\left(\frac{\dot{\phi}}{N}\right)^2L_{YZ} + 2(B^2-2E^2)E\left(\frac{\dot{\phi}}{N}\right)^2L_{ZW} \right] + 16\xi (H-\Sigma)^2E\frac{\dot{\phi}}{N}\,,
\end{eqnarray}
\begin{eqnarray}
 0 & = &
  \left\{
   -G_{3X}\left(\frac{\dot{\phi}}{N}\right)^2 - 4(H-\Sigma)\frac{\dot{\phi}}{N}\left[G_{4X}+4G_{4XX}\left(\frac{\dot{\phi}}{N}\right)^2\right] \right.\nonumber\\
 & & \left.- (H-\Sigma)^2 \left(\frac{\dot{\phi}}{N}\right)^2\left[3G_{5X} +G_{5XX}\left(\frac{\dot{\phi}}{N}\right)^2\right] - 8B^2\xi\chi^2 \right\}\frac{1}{N}\frac{d}{dt}\left(\frac{\dot{\phi}}{N}\right) \nonumber\\
 & &
  + 2\left[2G_4-2G_{4X}\left(\frac{\dot{\phi}}{N}\right)^2 - (H-\Sigma)\left(\frac{\dot{\phi}}{N}\right)^3G_{5X} + 4B^2\xi\chi^2\right]\left(\frac{\dot{H}}{N}-\frac{\dot{\Sigma}}{N}\right)
  \nonumber\\
 & & 
+ L - E^2\chi^2L_W - 4BE\chi^2L_Y - 2E^2\chi^2\left(\frac{\dot{\phi}}{N}\right)^2L_Z + 6(H-\Sigma)^2\left[G_4-G_{4X}\left(\frac{\dot{\phi}}{N}\right)^2\right]
  \nonumber\\
 & & - 2(H-\Sigma)^3\left(\frac{\dot{\phi}}{N}\right)^3G_{5X}
  - 4\xi\chi^2 \left[4B^2\left(H-\Sigma-\frac{\dot{\phi}}{N}\right)^2+ E^2(H-\Sigma)^2\right]\,, 
\end{eqnarray}
and
\begin{eqnarray}
 0 & = &
  -\left\{
    2(2L_{ZZ}\chi^2E^2+L_{XZ}) E\left(\frac{\dot{\phi}}{N}\right)^2
    + 2(4L_{YZ}B+L_{ZW}E)E^2\chi^2 + (4BL_{XY}+EL_{XW}+4EL_{Z})
  \right\}\chi^2\frac{\dot{\phi}}{N}\frac{\dot{E}}{N}
  \nonumber\\
 & & 
  +\left\{
   -L_X - 2E^2\chi^2L_Z - \left(\frac{\dot{\phi}}{N}\right)^2(L_{XX} + 4E^4\chi^4 L_{ZZ} + 4E^2\chi^2L_{XZ})\right.\nonumber\\
 & & \left.
      - 3H\frac{\dot{\phi}}{N}\left[2G_{3X}+G_{3XX}\left(\frac{\dot{\phi}}{N}\right)^2\right] 
      - 6(H^2-\Sigma^2)\left[ G_{4X} + 4G_{4XX}\left(\frac{\dot{\phi}}{N}\right)^2 + G_{4XXX}\left(\frac{\dot{\phi}}{N}\right)^4\right]\right.\nonumber\\
 & & \left.
      - (H+2\Sigma)(H-\Sigma)^2\frac{\dot{\phi}}{N}
      \left[6G_{5X}+7G_{5XX}\left(\frac{\dot{\phi}}{N}\right)^2+G_{5XXX}\left(\frac{\dot{\phi}}{N}\right)^4\right]\right\}\frac{1}{N}\frac{d}{dt}\left(\frac{\dot{\phi}}{N}\right)\nonumber\\
 & &
  - \left\{
     3G_{3X}\left(\frac{\dot{\phi}}{N}\right)^2 + 12H\frac{\dot{\phi}}{N}\left[G_{4X}+ G_{4XX}\left(\frac{\dot{\phi}}{N}\right)^2\right]  + 3(H^2-\Sigma^2)\left(\frac{\dot{\phi}}{N}\right)^2\left[3G_{5X}+G_{5XX}\left(\frac{\dot{\phi}}{N}\right)^2\right]
      + 8\xi B^2\chi^2\right\}\frac{\dot{H}}{N} \nonumber\\
 & & 
  + \left\{ 12\Sigma\frac{\dot{\phi}}{N}\left[G_{4X}+ G_{4XX}\left(\frac{\dot{\phi}}{N}\right)^2\right] + 6(H-\Sigma)\Sigma\left(\frac{\dot{\phi}}{N}\right)^2\left[3G_{5X}+G_{5XX}\left(\frac{\dot{\phi}}{N}\right)^2\right] - 16\xi B^2\chi^2\right\}\frac{\dot{\Sigma}}{N} \nonumber\\
 & &
  -3H\frac{\dot{\phi}}{N}L_X + (E^2-B^2)\chi^2L_W + 8BE\chi^2L_Y + 2E^2\frac{\dot{\phi}}{N}\chi^2\chi^2\left(H-4\Sigma-\frac{\dot{\phi}}{N}\right)L_Z \nonumber\\
 & &
  + \chi^2\frac{\dot{\phi}}{N}\left(2H-2\Sigma-\frac{\dot{\phi}}{N}\right)
  \Bigg\{
   4E^4\chi^2\left(\frac{\dot{\phi}}{N}\right)^2L_{ZZ}
   + (E^2-B^2)L_{XW} + 8EBL_{XY} + 2\left(\frac{\dot{\phi}}{N}\right)^2E^2L_{XZ}
		    \nonumber\\
 & & 
      + 16\chi^2E^3BL_{YZ} + 2E^2(E^2-B^2)\chi^2L_{ZW} \Bigg\}
 - 9\left(\frac{\dot{\phi}}{N}\right)^2H^2G_{3X}
 - 18\frac{\dot{\phi}}{N}H(H^2-\Sigma^2)\left[G_{4X}+\left(\frac{\dot{\phi}}{N}\right)^2G_{4XX}\right] \nonumber\\
 & &
  - 3\left(\frac{\dot{\phi}}{N}\right)^2H(H+2\Sigma)(H-\Sigma)^2
  \left[3G_{5X}+\left(\frac{\dot{\phi}}{N}\right)^2G_{5XX}\right]
+ 8\xi\chi^2\left[(H-\Sigma)^2E^2-(H+2\Sigma)^2B^2\right] \,,
\end{eqnarray}
where subscripts $X$, $W$, $Y$ and $Z$ represent derivatives with respect to them. Upon using
\begin{equation}
 \frac{\dot{\chi}}{N\chi} = - \left(2H-2\Sigma-\frac{\dot{\phi}}{N}\right)\,,
\end{equation}
it is easy to eliminate $\dot{\phi}$ in favor of $\dot{\chi}$ in the equations of motion.

\end{document}